\documentclass[11pt,a4paaper,oneside]{article}
\usepackage{lineno,hyperref}

\usepackage{graphicx}              
\usepackage{epsfig}
\usepackage{amsmath}               
\usepackage{amssymb}
\usepackage{epstopdf}
\usepackage{verbatim}
\usepackage{mathptmx}
\usepackage{mathalpha}
\usepackage{mathtools}
\usepackage[scale=0.80]{geometry}
\usepackage{graphicx, times}
\usepackage{amsfonts}              
\usepackage{amsthm}                
\usepackage{multicol}
\usepackage{algorithm}
\usepackage{algorithmic}
\usepackage{varwidth}
\usepackage{parskip}
\usepackage{hyperref}
\usepackage{rotating}
\usepackage[numbers,sort&compress]{natbib}
\usepackage{multirow}
\usepackage{pdflscape}
\usepackage[numbers]{natbib}
\usepackage{caption}
\usepackage{xcolor}
\usepackage{color}
\usepackage{comment}
\usepackage{subcaption}
\newcommand*{\rom}[1]{\expandafter\@\romannumeral #1}

\newcommand{\bea}{\begin{eqnarray}}
	\newcommand{\eea}{\end{eqnarray}}
\newcommand{\bee}{\begin{eqnarray*}}
	\newcommand{\eee}{\end{eqnarray*}}

\begin{document}
\author{Romanshu Garg$^{1}$\footnote{romanshugarg18@gmail.com}, Gyan Prakash Singh$^{1}$\footnote{gpsingh@mth.vnit.ac.in}, Ashutosh Singh$^{2}$\footnote{ashuverse@gmail.com}
\vspace{.3cm}\\
${}^{1}$ Department of Mathematics,\\ Visvesvaraya National Institute of Technology, \\ Nagpur 440010, Maharashtra, India.
\vspace{.3cm}\\
${}^{2}$ Centre for Cosmology, Astrophysics and Space Science (CCASS),\\
GLA University, Mathura 281406, Uttar Pradesh, India.}
\date{}

\title{Late-time dynamics of dark energy EoS in symmetric teleparallel gravity}

\maketitle

\begin{abstract}
In the symmetric teleparallel gravity framework, we study the cosmic dynamics of the universe with dark energy equation of state (EoS) parameter having non-linear forms. The non-metricity scalar induced by the dark energy EoS parameter evolves with time and, explains the physically reasonable transiting universe evolution in a consistent way. A comparative study has been presented to describe the ability of these models to fit the observational data. By using the Bayesian methods, we constrain the model parameters with the supernovae Ia (SneIa) and expansion rate data. We show that the expansion rate solutions may consistently describe the universe evolution based on cosmological indicators such as the effective EoS parameter, energy density, pressure, current age and the statefinder diagnostic. One may either have the quintom scenario or the future deceleration in these models subjected to the observational constraints.

\end{abstract}

\section{Introduction}\label{sec:1}
The confirmation from astronomical observations regarding the growing rate of expanding cosmos is an interesting finding of the observable universe \cite{1998AJ....116.1009R,1999ApJ...517..565P,2020A&A...641A...6P}. The problem emerges when it comes to the theoretical explanation of this observed phenomena. The General relativity (GR) is generally accepted as the most efficient gravitational theory to explain the different phenomena including the large-scale structure of the universe \cite{copeland2006dynamics}.
\par In the context of cosmology, the homogeneous and isotropic spacetime with the matter source provide interesting solutions for the scale factor $a(t)$, assisting in our comprehension of the universe's expansion. The universe must undergo two different accelerated expansion phases in order to recover the isotropy at late-times. One is the early-time inflation, whose dynamics may be studied  by introducing the scalar field in the Einstein-Hilbert action of GR and, second is the late-time accelerating expansion explained by a cosmological constant appearing in the Einstein's field equation. This model is the simplest model (which is also known as the $\Lambda$CDM model) and it fits very well with the known observations. Unfortunately, the cosmological tensions, cosmological constant and fine-tuning problems becomes the significant obstacles in success of this model  \cite{copeland2006dynamics,di2021realm,weinberg1989cosmological}. As a consequence, the critical examination of different approaches and/or generalizations of GR becomes somehow compulsory.  
\par In the recent decades, a number of techniques have been put up to deal with  these current cosmological issues. These days, the most popular option to solve the existing problems are rooted with the modified theory of gravity approach. The modification of General Relativity known as the $f(R)$ gravity \cite{buchdahl1970non} is one of the most well-known ideas to deal with the dark content problem of the cosmos, where $R$ is the Ricci Scalar. With the evolving time, several other modified theories are also widely explored \cite{harko2011f,elizalde2010lambdacdm,
bamba2010finite,asepjc,harko2010f,
hulke2020variable, Garg2023cfn,mandal2023cosmic,lalke2023late, singh2023cosmological, singh2022cosmological, singh2015bianchi,chaubey2016general, archanadixit2022,mandal2024late,as2024lyra,
SINGH2024865,asgrg,doi:10.1142/S0219887822501079,doi:10.1142/S0217751X23501695,as2021,nojiri2011unified, nojiri2017modified, capozziello2023role,
singh505abc,singh2020study,
johri1995gravitational,singh2001higher,dcmb}. 
\par In this direction, Jimenez et al. \cite{jimenez2018coincident} introduced a novel approach by considering a variation of the symmetric teleparallel equivalence to GR and it is known as the $f(Q)$ gravity, here $Q$ is a non-metricity scalar. The main geometric element that defines the characteristics of a gravitational interaction is the non-metricity $Q$, of the metric, which geometrically characterizes the fluctuation in the length of a vector in parallel transport. According to a number of investigations \cite{di2021realm,yang20212021}, the $f(Q)$ theory is one of the interesting alternative gravity interpretations for describing cosmological evolution of universe with the observations. Lazkoz et al. \cite{lazkoz2019observational} investigated the universe accelerated expansion to test compatibility with the observational data. Esposito et al.\cite{esposito2022reconstructing} examined the precise cosmological solutions that are isotropic as well as anisotropic. Harko et al. \cite{harko2018coupling} establish a class of $f(Q)$ theories in which $Q$ is connected non-minimally to the matter Lagrangian. Numerous further investigations have been completed in relation to the $f(Q)$ gravity theory\cite{jimenez2018coincident,yang20212021, lazkoz2019observational, esposito2022reconstructing, harko2018coupling, maurya2023transit, pradhan2022quintessence,capozziello2024preserving,nojiri2024well, hu2023nonpropagating,capozziello2022model,dimakis2022flrw,paliathanasis2024dipole, heisenberg2024review, heisenberg2024cosmological,rana2024phase, capozziello2022comparing,
ghosh2024dynamical,dcma,dimakis2021quantum,nm1,
nm2,nm3}. Capozziello et. al. have presented an approach to analyze the existence and the polarization of GWs in $f(Q)$ non-metric gravity by means of the geodesic deviation equation\cite{capozziello2024gravitational}. Capozziello et. al. are investigate the polarization modes of gravitational waves in $f(Q)$ non-metric gravity without gauge fixing \cite{capozziello2025gravita}. Capozziello et. al. discussed the existence, nature and properties of the polarization modes of GWs in $f(Q, B)$ nonmetric gravity both in coincident gauge, turning off the STG connection, and in free gauge, without gauge fixing\cite{capozziello24}.
\par In the present paper, we explore an analytical solution based cosmological scenario to find whether $f(Q)$ gravity can be modeled based on varying dark energy EoS parameter having non-linear form or not?. The EoS parameters having non-linear nature may be solved to explore the nature of non-metricity scalar. In particular, we consider a class of parametrization for dark energy usually termed as the ``Gong-Zhang parametrization(s)" \cite{gong2005probing,ChinPhysLett.40.019801,CASTILLOSANTOS2023101225}. These non-linear parametrizations are given by $\omega_{DE}(z)=\frac{w_{0}}{(1+z)}$ and  $\omega_{DE}(z)=\frac{w_{0}}{(1+z)} \exp\left(\frac{z}{1+z}\right)$. These forms involve one model parameter and, the EoS parameter for dark energy do not diverge at either at the early stage or at future during the universe evolution. In the model independent approaches, the parametrizations of dark energy equation of state parameters have been often used to study the dynamic evolution of universe in models. These parameters may describe the accelerating universe expansion at late stages of universe evolution. The considered parametrizations having one parameter may enable us to study the geometry of universe in simplified way with other density parameters \cite{gong2005probing}. One may visualize the dynamic dark energy evolution in comparison to cosmological constant model. In the Gong-Zhang models, the dark energy may track the matter in past. These models may even have the past of matter-dominated evolution. These characteristics may induce quite important implications in the late universe evolution and may also be of phenomenological interest. We probe whether the induced cosmic dynamics may describe a consistent description of universe in $f(Q)$ gravity, subjected to the SneIa and expansion rate data or not? The behavior of non-metricity in these models may describe the evolution of world-lines and thus telling us the geometry of the universe. We examine these EoS parameters in the context of $f(Q)$ cosmological model using data analytic approaches. 
\par This work has been summarized into six sections, which are as follows: In section (\ref{sec:2}), we discuss the general $f(Q)$ theory in the flat FLRW metric. In section (\ref{sec:3}), we derive the non-metricity scalar of the power law $f(Q)$ gravity with the non-linear forms of EoS parameters. In section (\ref{sec:5}), we constrains the model parameter by employing the Bayesian $\chi^2$ approach. In section (\ref{sec:6}), the cosmological properties have been studied, subjected to the constrained parameters. A detailed review of the results are provided in Sec. (\ref{sec:7}).

\section{The basic equations in $ f(Q)$ gravity}
\label{sec:2}
An overview of the $f(Q)$ gravity will be provided in the present section. The $f(Q)$ gravity action may be provided by \cite{jimenez2018coincident}.
\begin{equation}{\label{1}}
S = \int \left[-\frac{1}{2}f(Q) + \mathcal{L}_m\right] \sqrt{-g} \, d^4x,
\end{equation}
where arbitrary function of nonmetricity scalar $Q$ can be represented by $f$ and the matter Lagrangian density is denoted by $\mathcal{L}_m$. Here, $g$ represents the determinant of metric tensor $g_{\mu \nu}$. There are two distinct traces of $Q_{\alpha \mu \nu}$:
\begin{equation}{\label{2}}
Q_\sigma = Q_\sigma^{\phantom{\sigma}\mu} \mu_{\phantom{\mu}}, \quad \tilde{Q}_\sigma = Q^{\mu\phantom{\mu}}_{\phantom{\mu}\sigma \mu}.
\end{equation}
Moreover, the non-metricity scalar is represented as a contraction of  $Q_{\alpha \mu \nu}$ which is given by
\begin{equation}{\label{3}}
Q=-Q_{\sigma\mu\nu} P^{\sigma\mu\nu},
\end{equation}
where $P^{\sigma\mu\nu}$ is the superpotential tensor (also known as the non-metricity conjugate) and it follows
\begin{equation}{\label{4}}
4P^\sigma_{\phantom{\sigma}\mu\nu} = -Q^\sigma_{\phantom{\sigma}\mu\nu} + 2Q_{(\mu\phantom{\sigma}\nu)}^{\phantom{\mu}\sigma} - Q^\sigma g_{\mu\nu} - \tilde{Q}^\sigma g_{\mu\nu} - \delta^\sigma_{(\mu} Q_{\nu)}^{\phantom{\nu)}},
\end{equation}
By varying the action (\ref{1}) with respect to the metric, the field equations can be determined as
\begin{equation}{\label{5}}
\frac{2}{\sqrt{-g}} \nabla_\sigma \left( \sqrt{-g} f_Q P^{\sigma}_{\mu\nu} \right) + \frac{1}{2} g_{\mu\nu}f + f_Q \left( P_{\mu\sigma\beta} Q^{\sigma\beta}_{\phantom{\sigma\beta}\nu} - 2Q_{\sigma\beta \mu} P^{\sigma\beta}_{\phantom{\sigma\beta}\nu} \right) = T_{\mu\nu},
\end{equation}
where $f_Q = \frac{df(Q)}{dQ}$ and $
T_{\mu\nu}=-\frac{2}{\sqrt{-g}} \frac{\delta(\sqrt{-g} \mathcal{L}_m)}{\delta g^{\mu\nu}}$ with a choice of units such that $8\pi G = c = 1$.\\
The energy-momentum tensor $T_{\mu\nu}$ may also be provided by $ T_{\mu\nu} =  p g_{\mu\nu}+(p+\rho) u_\mu u_\nu ,$ here the isotropic pressure and energy density of a perfect fluid are symbolized by $p$ and $\rho$, respectively. $u_\mu$ is the $4$-velocity and it satisfies the normalization restriction $u_\mu u^\mu = -1$.\\
Throughout the study, we will consider a spatially flat FLRW universe \cite{partridge2004introduction} whose metric is provided by
\begin{equation}{\label{7}}  
ds^{2}=a^{2}(t) \left( dx^{2}+ dy^{2}+ dz^{2}\right)-dt^{2},
\end{equation}
In this framework, the non-metricity scalar $Q$ may be expressed as $Q=6H^{2}$, where $H\equiv \frac{\dot{a}}{a}$ is representing the Hubble parameter and  $a(t)$ is the scale factor. The dot symbolizes derivative with regards to cosmic time $t$. For the metric (\ref{7}), the corresponding Friedmann equations  are \cite{jimenez2020cosmology}: 
\begin{eqnarray}
3H^2 = \frac{1}{2f_Q} \left( \rho + \frac{f}{2} \right), \quad \dot{H} + H \left( 3H + \frac{\dot{f}_Q}{f_Q} \right) = \frac{1}{2f_Q} \left( \frac{f}{2}-p  \right)\label{9}
\end{eqnarray}
The standard Friedmann equations of GR may be obtained if we select the function $f(Q)$ as $f(Q)=Q$ \cite{jimenez2020cosmology}. We proceed with $f(Q)=Q+F(Q)$ and thus the field Eqs. (\ref{9}) can be described as follows 
\begin{eqnarray}
3H^2 =  \frac{F}{2} - Q F_{Q} +\rho, \quad \dot{H}(F_{Q} +2QF_{QQ}+ 1)+\frac{1}{4} \left(- F + Q + 2QF_Q \right) = \frac{-p}{2} \label{11}
\end{eqnarray}
The Eqs. (\ref{11}) can be viewed as the analogs of GR cosmology, incorporating an additional term due to the non-metricity $Q$ of space-time. This term may exhibit the behavior of a fluid component associated with the dark energy, i.e. $\rho_{Q}=\rho_{DE}$ and $p_{Q}=p_{DE}$. Consequently, through Eqs. (\ref{11}), we get
\begin{eqnarray}
3H^2=(\rho+\rho_{DE}), \quad 2\dot{H}+3H^2= -(p+p_{DE}), \label{13}
\end{eqnarray}
here $p_{DE}$ and $\rho_{DE}$ represent the pressure and density contributions to the DE as caused by of space-time's non-metricity which is defined by
\begin{equation}{\label{14}}
\rho_{DE} = \frac{F}{2} -QF_{Q}, \quad p_{DE} = 2\dot{H}(2QF_{QQ} + F_Q) - \rho_{DE}.
\end{equation}
Moreover, because of the DE component, the equation of state (EoS) parameter is
\begin{equation}{\label{16}}
\omega_{DE}=\frac{p_{DE}}{\rho_{DE}}=-1+\frac{4\dot{H} (2QF_{QQ} + F_Q)}{F - 2QF_{Q}}. 
\end{equation}

\section{Cosmological model with non-linear EoS Parameter}
\label{sec:3}
For the purpose of investigation of Gong-zhang  parametrized form of EoS parameter for DE, we assume the functional form $F(Q)=\alpha Q^{n}$, where $n$ and $\alpha$ are model parameters. Khyllep et al. \cite{khyllep2021cosmological} provides the motivation for this specific functional form of $f(Q)$. When $n=1$, this scenario corresponds to the symmetric teleparallel equivalent of general relativity, with Newton's gravitational constant re-scaled by a factor of $(\alpha +1)$ \cite{jimenez2020cosmology}. \\
We obtained the pressure $p_{DE}$ and energy density $\rho_{DE}$ for DE in terms of the $H$ using this form of $F(Q)$ as
\begin{eqnarray}
\rho_{DE}=\alpha {6}^{n}\left(\frac{1}{2}-n\right)H^{2n}, \quad p_{DE}=-2\alpha 6^{n-1}\left(\frac{1}{2}-n \right)H^{2(n-1)} (3H^2 + 2n\dot{H}), \label{18}
\end{eqnarray}
For the dark energy, its EoS parameter $\omega_{DE}$ would become $\omega_{DE}= -1- \frac{2n}{3}\left(\frac{\dot{H}}{H^2} \right)$. By using relation $ \dot{H}=-H(1+z)\frac{dH}{dz} $, we get 
\begin{equation}{\label{20}}
\omega_{DE}= -1+\frac{2n(1+z)}{3H}\left(\frac{dH}{dz}\right),
\end{equation}
The scale factor $a(t)$, $\rho_{DE}$ and $p_{DE}$ are three unknowns quantities. While we have independent equations (\ref{18}) and thus to solve equation (\ref{20}), one more equation is required. To resolve this, we adopt two distinct parametric forms of the equation of state parameter for dark energy. We consider the Gong-zhang parametrizations of the EoS parameter for DE in terms of red-shift $z$ \cite{gong2005probing}.
\subsection{Model I}\label{sec:3.1}
In this model, we proceed with \cite{gong2005probing}
\begin{equation}{\label{21}}
\omega_{DE}(z)=w_0(1+z)^{-1} 
\end{equation}
This EoS parameter for dark energy consists one parameter $w_0$ and may be constrained using the observations. In framework of cosmological modeling, the EoS of dark energy (DE) parametrization may provide a number of significant advantages. First of all, it offers a framework that can be physically understood with model parameter $w_{0}$ denoting the DE EoS parameter at the current epoch $(z=0)$. By explicitly incorporating the red-shift dependence into the parametrization above, we may be able to represent how DE evolves across time. When $z$ approaches closer to infinity (in the past) $\omega_{DE}(z) \sim  0$, that shows the DE EoS parameter tends toward zero in the early era. As $z \to -1$ (in the asymptotic limits) $\omega_{DE}(z) \to - \infty $,  suggests that the DE EoS parameter tends toward negative infinity in the distant future. These features are different from the cosmological constant kind of dark energy where $\omega=-1$. These features of the DE EoS parameter are reasonably interesting, since $w_0$ is close to $-1$ in observational findings \cite{2020A&A...641A...6P}. We probe the resulting dynamics in in $f(Q)$ gravity. 
\vspace{0.2cm} \\
From Eqs. (\ref{20}) and (\ref{21}), one may get
\begin{equation}{\label{22}}
-1+\frac{2n(1+z)}{3H}\left(\frac{dH}{dz} \right)=w_{0}(1+z)^{-1}
\end{equation}
Solving equation (\ref{22}),  we determine the Hubble parameter in relation to red-shift 
\begin{equation}{\label{23}}
H(z)=h_{0}\exp\left(\frac{3w_{0}z}{2n(1+z)}\right)(1+z)^{\frac{3}{2n}},
\end{equation}
The Hubble parameter's current value is represented by $h_{0}$ in the model. The model possess three parameters $h_0,n$ and $w_0$.
\subsection{Model II}\label{sec:3.2}
In this model, we proceed with \cite{gong2005probing} 
\begin{equation}{\label{101}}
\omega_{DE}(z)=\frac{w_{0}}{(1+z)} \exp\left(\frac{z}{1+z}\right)
\end{equation}
where $w_0$ is model parameter denoting the present day value of dark energy in model. In the asymptotic past, when $z \gg 1$, $ \omega_{DE} \sim 0 $. In the asymptotic future, for $z \to -1$ , $ \omega_{DE} \to 0$. This is the main distinction between this model (\ref{101}) with model (\ref{21}). In the past, these two forms behaved nearly identically but they behave significantly differently in the future.
\vspace{0.2cm}\\ 
From equation (\ref{20}) and equation (\ref{101})
\begin{equation}{\label{102}}
-1+\frac{2n(1+z)}{3H}\left(\frac{dH}{dz} \right)=\frac{w_{0}}{(1+z)} \exp\left(\frac{z}{1+z}\right)
\end{equation}
Solving equation (\ref{102}), we determine the following expression of the Hubble parameter with respect to red-shift
\begin{equation}{\label{103}}
H(z)=h_{0}\exp\left(\frac{-3w_{0}}{2n}\left[1-\exp  \left( \frac{z}{1+z}\right)  \right]\right)(1+z)^{\frac{3}{2n}},
\end{equation}
where $h_0$ is the Hubble parameter value at $z=0$. The model possess three parameters $h_0,n$ and $w_0$.

\section{Observational constraints on model parameters}
\label{sec:5}
In this section, we proceed to conduct a Bayesian study of the present cosmological model for its observational compatibility with the cosmic chronometer and supernovae type Ia data. To find the model parameter's constraint values $h_{0}$, $w_{0}$ and $n$, we employ two observational datasets of late-times such as the cosmic chronometer (CC) sample and the joint data sample which is a combination of CC and Pantheon nomenclated as the CC+Pantheon sample. We use the $\chi^{2}$ minimization technique with the Markov Chain Monte Carlo (MCMC) approach implemented in the emcee Python library~\cite{foreman2013emcee}.

\subsection{The Cosmic chronometer data}
\label{sec:5.1}
\begin{center}
\begin{figure}[t]
\centerline{\includegraphics[width=12.8cm,height=8cm]{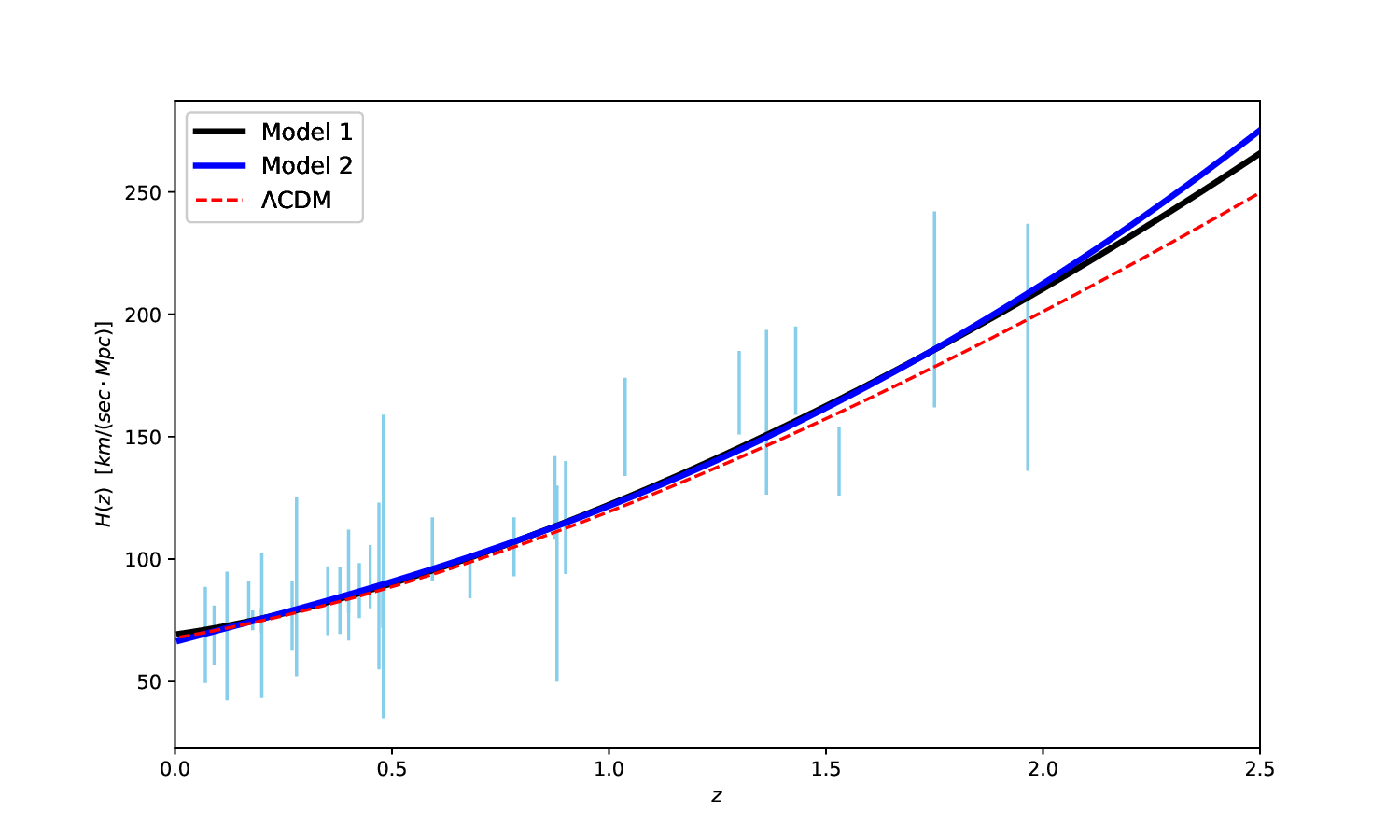}}
\caption{In comparison to the $\Lambda CDM$ model, the best fit Hubble parameters  (given by Eq. (\ref{23}) and (\ref{103})) versus $\mathit{z} $.}
\label{fig:1}
\end{figure}
\end{center}

In this part, to constrain the model parameters values, we use Hubble parameter measurements determined via the various age approach from the red-shift range $0.07 \leq z \leq 1.965$ \cite{stern2010cosmic,moresco2015raising} often known as CC data. In this case, we consider this data set having 31 data points \cite{simon2005constraints,sharov2018predictions}. Jimenez and Loeb \cite{jimenez2002constraining} proposed a basic principle for the cosmic chronometer observations that relates the red-shift $(z)$, cosmic time $(t)$, and the Hubble parameter $H(z)$ as follows: $H(z)=\frac{-1}{(1+z)}\frac{dz}{dt}$. For the observational constraints on the model parameters $h_{0}, w_{0}$, and $n$, we minimize the $\chi^{2}$ function (which is equivalent to maximizing the likelihood function) expressed as \cite{mandal2023cosmic,mandal2024late,  SINGH2024100827,as2022,aspdu}
\begin{equation}{\label{24}}
\chi^{2}_{CC}(\theta)=\sum_{i=1}^{31} \frac{[H_{th}(\theta,z_{i})-H_{obs}(z_{i})]^{2}    }{ \sigma^{2}_{H(z_{i})}}.   
\end{equation}
Here, $H_{th}$ indicates the Hubble's theoretical value, $H_{obs}$ symbolizes the Hubble observed value, standard error of the observed value is represents by $\sigma_{H}$.\\ 
The CC data is composed of $31$ uncorrelated points of Hubble parameter observations from the passively evolving galaxies \cite{simon2005constraints}. We use the observed data points with their errors along with the theoretical Hubble parameters in model I (\ref{23}) and II (\ref{103}) to get constraints on model parameters using $\chi^{2}_{CC}$ definition. By using the constraints obtained from MCMC analysis, the best fit $H(z)$ curves of the considered models are compared with the corresponding $H(z)$ of $\Lambda$CDM model in Fig. (\ref{fig:1}). In the $\Lambda$CDM model, we use $\Omega_{m0}=0.303,H_0=67.4 \ km/(sec\cdot Mpc)$.  The universe in models traces its evolution from the decelerating phase into the accelerating phase. The behaviors of corresponding Hubble parameters of models reveal that the evolution dynamics may be different in the past. However, in the recent past up to $z\approx 1.2 $, these models are compatible with the $\Lambda$CDM model. 


\subsection{The Pantheon data}\label{sec:5.2}
\begin{center}
\begin{figure}
\includegraphics[width=19.8cm,height=17.9cm]{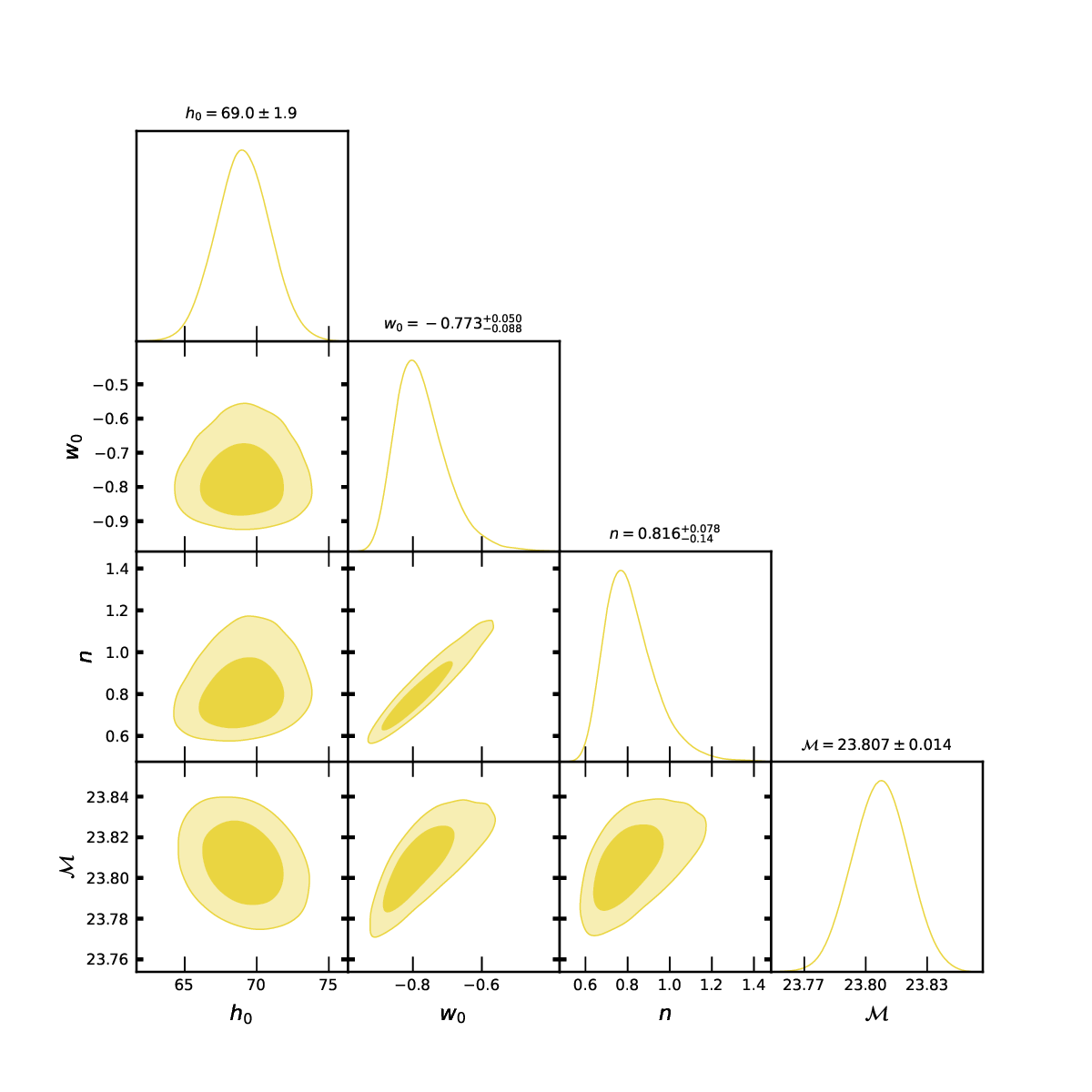}
\caption{Model I: Marginalized 1D and 2D contour maps with median values of $h_{0}$, $w_{0}$ and $n$ using Joint data set.}
\label{fig:5}
\end{figure}
\end{center}

\begin{center}
\begin{figure}
\includegraphics[width=19.8cm,height=19cm]{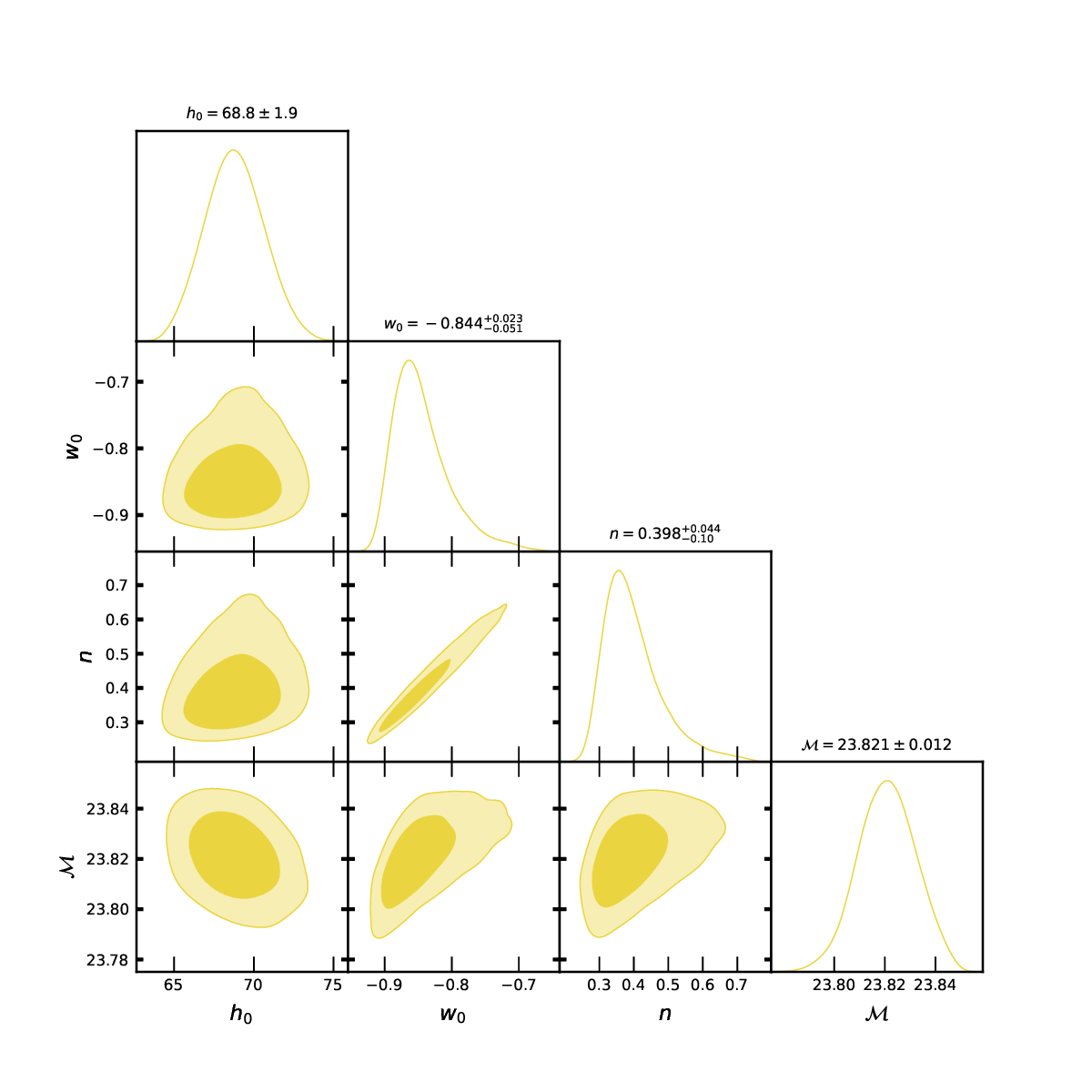}
\caption{Model II: Marginalized 1D and 2D contour maps with median values of $h_{0}$, $w_{0}$ and $n$ using Joint data set.}
\label{fig:6}
\end{figure}
\end{center}

We employ the Pantheon sample that contains 1048 supernovae Type Ia (SNIa) data points for the red-shift range $0.01 < z < 2.26$ ~\cite{scolnic2018complete}. The SNIa sample includes contributions from the SNLS~\cite{guy2010supernova},The CfA1-CfA4~\cite{riess1999bvri,hicken2009improved} surveys, SDSS~\cite{sako2018data},Pan-STARRS1 Medium Deep Survey~\cite{scolnic2018complete}, Carnegie Supernova Project (CSP)~\cite{contreras2010carnegie}. \\
In the MCMC analysis using Pantheon dataset, we use the theoretically expected apparent magnitude $m_{th} (z)$ defined by 
\begin{equation}{\label{7a}}
m_{th}(z)=M+5\log_{10}\left[\frac{d_{L}(z)}{Mpc}\right]+25,
\end{equation}
where $M$ is absolute magnitude. Moreover, the luminosity distance $d_{L}(z)$ (having dimension of the Length) can be described as ~\cite{mandal2024late,SINGH2024100827,as2022,aspdu}
\begin{equation}{\label{8a}}
d_{L}(z)=c(1+z)\int_{0}^{z}\frac{dz'}{H(z')},
\end{equation}
where $c$ is the speed of light and $z$ is the red-shift of SNIa as measured in the cosmic microwave background (CMB) rest frame. The luminosity distance $(d_{L})$ is often substituted with the Hubble-free luminosity distance $(D_{L}(z) \equiv H_{0}d_{L}(z)/c)$. The equation (\ref{7a}) would also be rewritten as
\begin{equation}{\label{9a}}
m_{th}(z)=M+5\log_{10}\left[D_{L}(z)\right]+5\log_{10}\left[\frac{c/H_{0}}{Mpc}\right]+25. 
\end{equation}
where $\mathcal{M}$ is defined by combining the parameters $M$ and $H_{0}$ as
\begin{equation}{\label{10a}}
\mathcal{M}\equiv M+5log_{10} \left[\frac{c/H_{0}}{Mpc}\right]+25=M-5\log_{10}(h)+42.38, 
\end{equation} 
where $H_{0}=h \times 100$ Km/(s.Mpc). For Pantheon data, we utilize this parameter together with relevant $\chi^{2}$ in the MCMC analysis as~\cite{mandal2024late,SINGH2024100827,as2022,asvesta2022observational}
\begin{equation}{\label{11a}}
\chi^{2}_{P}= \nabla m_{i}C^{-1}_{ij}\nabla m_{j},
\end{equation}
where $\nabla m_{i}=m_{obs}(z_{i})-m_{th}(z_{i})$, $C_{ij}^{-1}$ is covariance matrix's inverse \cite{asvesta2022observational} and, $m_{th}$ will be determine by equation (\ref{9a}). The luminosity distance depends on the Hubble parameter. Consequently, we utilize equations (\ref{23}), (\ref{103}) and the emcee package~\cite{foreman2013emcee} to obtain the maximum likelihood estimate using the joint (CC+Pantheon) dataset. For the maximum likelihood estimate, the joint $\chi^{2}$ is expressed as $\chi^{2}_{CC}+\chi^{2}_{P}$. We use the joint $\chi^{2}$ definition for the CC and Pantheon data in MCMC analysis to plot the posterior distributions of parameters. We use the uniform priors for model parameters. For the joint data set, we use $48$ random chains (walkers) and 12000 iterations (steps) in the MCMC analysis. The Figs. (\ref{fig:5}) and (\ref{fig:6}) are displaying the marginalized 1D and 2D contour map distribution obtained from the Markov Chain Monte Carlo analysis using joint (CC+Pantheon) data. \\
The median values of the model parameters (obtained from \textit{emcee} using the MCMC analysis) for Hubble parameters of Model I (\ref{23}) and Model II (\ref{103}) are summarized in Table (\ref{table:1}) and (\ref{table:2}) respectively.
\begin{table}[htbp]
\centering
\begin{tabular}{|c|c|c|c|c|c|c|c|}
\hline
Dataset & $h_{0}$ [Km/(s.Mpc)] & $w_{0}$  & $n$ & $\mathcal{M}$ & $q_{0}$ & $z_{t}$ & $t_{0}$ \\
\hline
CC & $69.0^{+1.4}_{-1.4}$ & $-0.807^{+0.057}_{-0.077}$ &$0.756^{+0.050}_{-0.069}$ & - & -0.6170 & 0.628 & $13.17^{+0.30}_{-0.19}$ \\
\hline
joint  & $69.0^{+1.9}_{-1.9}$  & $-0.773^{+0.050}_{-0.088}$  &  $0.816^{+0.078}_{-0.14}$ & $23.807^{+0.014}_{-0.014}$ & -0.5827 & 0.700 & $13.54^{+0.02}_{-0.43}$\\
\hline
\end{tabular}
\caption{ {\bf{Model I}}: Model parameters's median values with the present values of $q_{0}, z_{t}, t_{0}$.}
\label{table:1}
\end{table}

\begin{table}[htbp]
\centering
\begin{tabular}{|c|c|c|c|c|c|c|c|}
\hline
Dataset & $h_{0}$ [Km/(s.Mpc)] & $w_{0}$  & $n$ & $\mathcal{M}$ & $q_{0}$ & $z_{t}$ & $t_{0}$ \\
\hline
CC & $66.00^{+0.90}_{-1.0}$ & $-0.756^{+0.037}_{-0.056}$ &$0.494^{+0.047}_{-0.074}$ & - & -0.2576 & 0.7 & $12.77^{+0.10}_{-0.10}$ \\
\hline
joint  & $68.8^{+1.9}_{-1.9}$  & $-0.844^{+0.023}_{-0.051}$  &  $0.398^{+0.044}_{-0.100}$ & $23.821^{+0.012}_{-0.012}$ & -0.4120 & 0.785 & $12.45^{+0.07}_{-0.31}$\\
\hline
\end{tabular}
\caption{ {\bf{Model II}}: Model parameters's median values with the present values of $q_{0}, z_{t}, t_{0}$.}
\label{table:2}
\end{table}

\section{Physical and dynamical properties of the models}
\label{sec:6}
In this section, we investigate the dynamical characteristics of the universe evolution in model I and II by using the observationally constrained values of model parameters. 
\subsection{Deceleration parameter}\label{sec:6.1}
One of the important parameter to characterize the dynamics of Universe expansion phase is the deceleration parameter $(q)$. One may explore the cosmos's expansion history depending on the $q$ values as: $q>0$ for the decelerated phase, $q<0$ for the accelerated phase and $-1 < q < 0$ for power-law accelerated expansion. For $q<-1 \ (q=-1)$, the universe exhibits a super-exponential (de Sitter) expanding epoch respectively \cite{as2024lyra,SINGH2024865,doi:10.1142/S0219887822501079,doi:10.1142/S0217751X23501695,as2021}. This cosmographic parameter may be expressed as \begin{equation}{\label{29}}
 q = -1 + \frac{d}{dt}\frac{1}{H}.
\end{equation}
By using (\ref{23}), (\ref{103}) and (\ref{29}), we may obtain
\begin{eqnarray}
q(z)=-1+{\frac{3[w_{0}+z+1]}{2n(1+z)}} \label{30} \quad (\text{Model I}) \\
q(z)=-1+ {\frac{3\left[ z+1+w_{0}\exp \left(\frac{z}{1+z}\right) \right]}{2n(1+z)}} \label{104}  \quad (\text{Model II})
\end{eqnarray}

\begin{figure}[!htb]
\captionsetup{skip=0.4\baselineskip,size=footnotesize}
   \begin{minipage}{0.40\textwidth}
     \centering
     \includegraphics[width=9.0 cm,height=7.5cm]{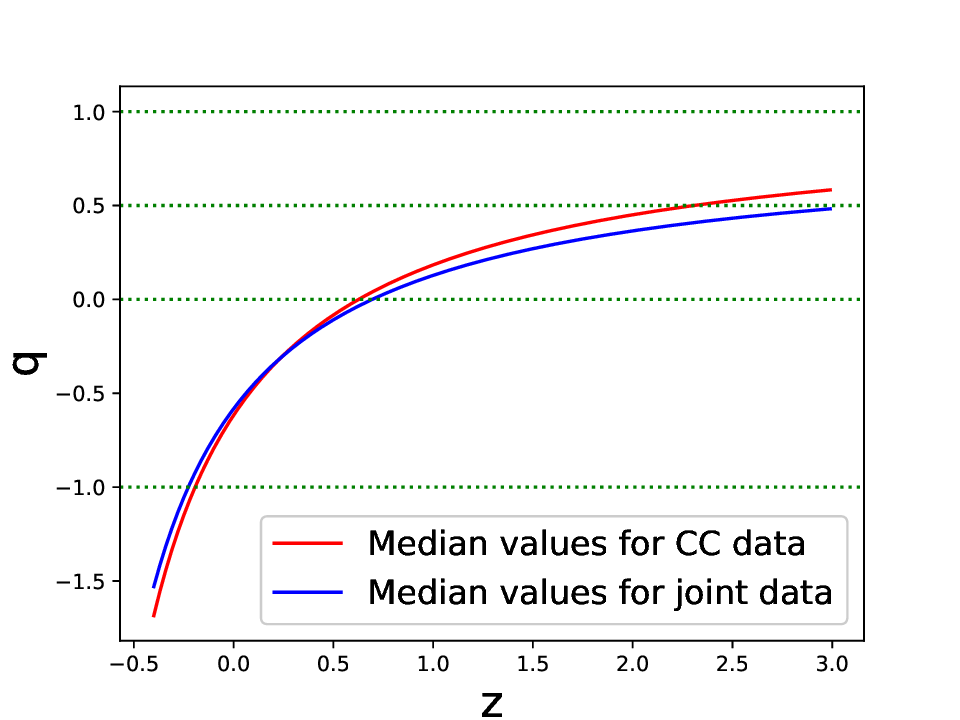}
\caption{Model I: The deceleration parameter versus $z$.}
\label{fig:7}
    \end{minipage}\hfill
   \begin{minipage}{0.40\textwidth}
     \centering
     \includegraphics[width=9.0 cm,height=7.5cm]{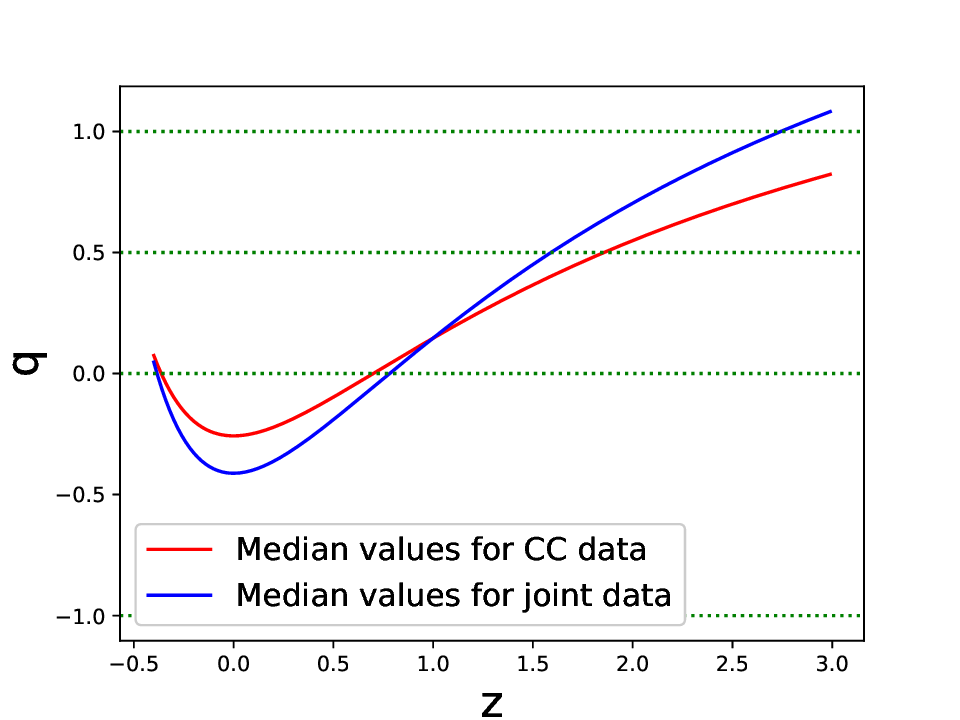}
  \caption{Model II: The deceleration parameter versus $z$.}
\label{fig:8}
   \end{minipage}
\end{figure}
The deceleration parameter (\ref{30}) of model I may explain the deceleration-acceleration transit of the universe. For median values, the behavior of $q$ is depicted in Figure $(\ref{fig:7})$. At $z=0.628$ ($z=0.7$), the deceleration parameter is zero for the constrained values from CC (Joint) data respectively. At present, the deceleration parameter values are close to $q_0\approx -0.6$. In particular, $q_{0}=-0.617$ and $q_{0}=-0.582$ for CC and Joint estimates respectively. This negative value indicates that the universe's expansion is accelerating at present $(z=0)$. The trajectory of $q$ for the joint estimates of model parameter, $q=\frac{1}{2}$, which highlights the matter dominated past in the model. However, for the CC estimates of the model parameters, the model may possess matter having EoS parameter greater than zero. It simply mean that the dominated component in the past may not be the cold dark matter in this model. The behavior of $q$ suggest that the universe is accelerating due to the dominating component having EoS parameter less than $-\frac{1}{3}$. But, in future, the universe may accelerate with super-exponential rate under the influence of phantom kind of dark energy. Since, the accelerated expansion of the universe crosses the de Sitter line, we may term this kind of dark energy as the quintom dark energy \cite{zhao2006quintom}. \\
The universe transitions from a decelerated expansion phase into an accelerated expansion phase in model II and this transition may occur at $z=0.7$ and $z=0.785$ for CC and joint estimates respectively. And, the deceleration parameter's present value are $q_{0}=-0.2576$ and $q_{0}=-0.4120$ for CC and Joint estimates respectively. The behavior of $q$ suggest that the model may not have a cold dark matter dominated era. It means that the structure formation era may not be explained in the model. During present times, the model may explain the accelerated expansion era, which is compatible with the observations. But, it has sincere shortcoming especially during the past evolution. In future, this model predicts the deceleration era. There are several instances in literature, which predicts this kind of behavior in the future \cite{CHAKRABORTY2014424,CASTILLOSANTOS2023101225,PhysRevD.109.023514}.

\subsection{Energy Density and Pressure of DE}
\label{sec:6.2}

 The energy density continues to be positive throughout the expansion history, but the pressure may have become negative recently.  During the transition from deceleration to acceleration, due to the dominance of dark energy, the pressure become negative and the energy density stays positive.\\
Using equation (\ref{18}), (\ref{23}) and (\ref{103}), we determine the energy density and pressure for DE as   
\begin{equation}
\rho_{DE}(z)=\frac{\alpha}{2} 6^{n}(1-2n)h_{0}^{2n}(1+z)^{3}\exp\left(\frac{3zw_{0}}{1+z}\right) \quad \text{Model I} {\label{31}}  \\
\end{equation}

\begin{equation}
p_{DE}(z)=\frac{\alpha}{2} 6^{n} w_{0}(1-2n)h_{0}^{2n}(1+z)^{2}\exp\left(\frac{3zw_{0}}{1+z}\right) \quad \text{Model I}  \label{32} \\
\end{equation}
	
\begin{equation}	
	\rho_{DE}(z)=\frac{\alpha}{2} 6^{n}(1-2n) \left( h_{0}.\exp\left(\frac{-3w_{0}}{2n}\left[1-\exp  \left( \frac{z}{1+z}\right)  \right]\right)(1+z)^{\frac{3}{2n}}\right)^{2n} \quad \text{Model II}  \label{105} \\ 
\end{equation}

\begin{equation}	
p_{DE}(z)=\frac{\alpha 6^{n} w_{0}(1-2n) \exp\left(\frac{z}{1+z} \right)  \left( h_{0}.\exp\left(\frac{-3w_{0}}{2n}\left[1-\exp  \left( \frac{z}{1+z}\right)  \right]\right)(1+z)^{\frac{3}{2n}}\right)^{2n}}{2(1+z)} \quad \text{Model II} \label{106} 
\end{equation}
For the constrained values of the model parameters, the energy densities for DE ($\rho_{DE}$) in both models exhibit an increasing nature with red-shift $(z)$(which corresponds to decreasing nature over cosmic time $(t)$) and remain positive throughout the expansion. Energy density decreases as the universe transitions from early stages to later ones. The energy density for DE ($\rho_{DE}$) displays the expected positive behavior, emphasizing a contribution to the expansion of the Universe. The energy densities will have positive values rules out the existence of finite-time future singularities in the models. \\
Using the Hubble parameter expression, we obtained the dark energy density by eq. (12). Using right hand side of eq. (9), we may obtain the total energy density of the universe in the models. Since, $H$ and $\rho_{DE}$ are known, we may obtain the energy ($\rho$) and pressure $(p)$ of matter satisfying EoS parameter $\omega=0$. In the effective scenario, the energy density ($\rho$) and pressure $(p)$ of matter appearing in eq. (9) is responsible for the decelerating expansion of the past era.\\
In both models, the pressure for DE $(p_{DE})$ begins with negative values in the early universe (at high redshift) for median values of model parameters. The pressure continues to be negative during the current and subsequent the later epochs. The non-metricity scalar appears to be responsible for the negative pressure essential for the universe's accelerated expansion in the late era. \\  
The universe may be expanding more quickly in this late era due to the pressure's negative nature. These observations are consistent with the accelerating Universe's expanding nature. In model I, we take $\alpha=-1$ and in model II, we take $\alpha=1$ for these graphical illustrations. 

\subsection{Equation of State Parameter}\label{sec:6.3}
The EoS parameter represents the correlation between pressure $(\mathit{p})$ and energy density $(\rho)$. The EoS parameter illustrates several characteristics including the dust $(\omega=0)$, the radiation dominated period  $(\omega=\frac{1}{3})$, and the vacuum energy $(\omega=-1)$ corresponding to the de Sitter stage in $\Lambda$CDM model. Additionally, there is the accelerating period of the Universe, represented by $\omega <- \frac{1}{3}$. This period includes the phantom regime $(\omega < -1)$ and the quintessence regime $(-1 < \omega <- \frac{1}{3})$. 

\begin{figure}[!htb]
\captionsetup{skip=0.4\baselineskip,size=footnotesize}
   \begin{minipage}{0.40\textwidth}
     \centering
     \includegraphics[width=10.0 cm,height=7.5cm]{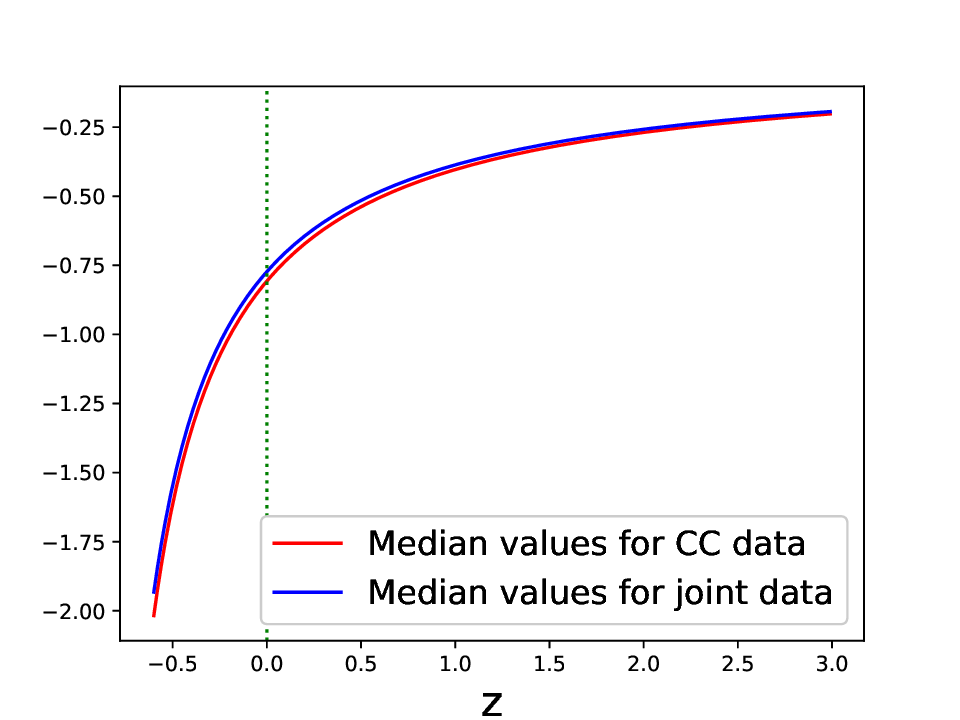}
\caption{Model I: $\omega_{DE}$ versus $\mathit{z}$}
\label{fig:13}
    \end{minipage}\hfill
   \begin{minipage}{0.40\textwidth}
     \centering
     \includegraphics[width=9.5 cm,height=7.5cm]{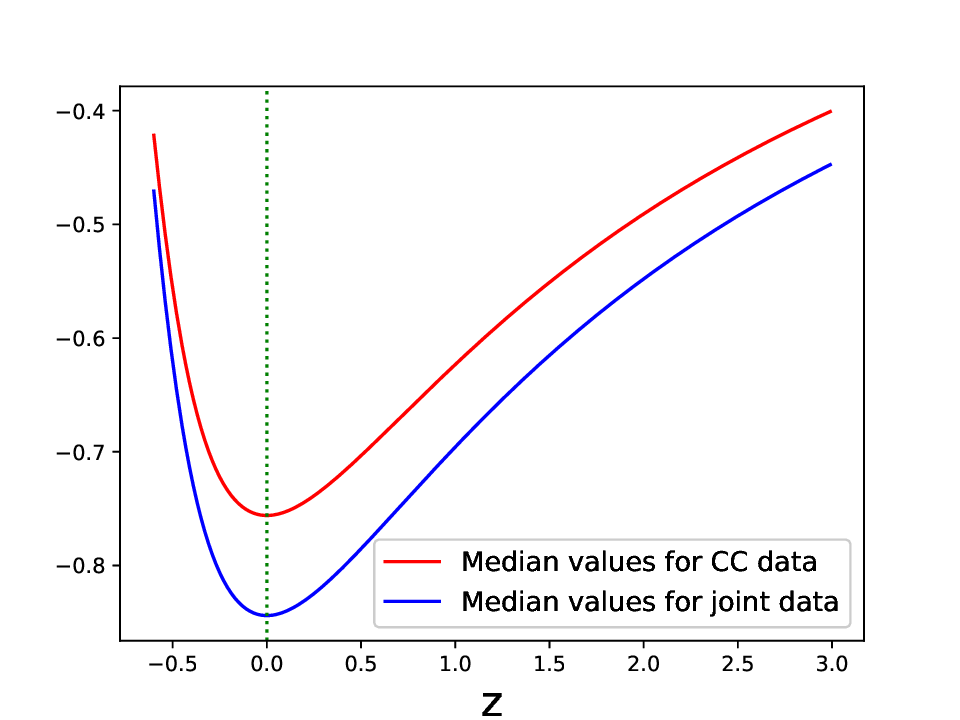}
  \caption{Model II: $\omega_{DE}$ versus $\mathit{z}$ }
\label{fig:14}
   \end{minipage}
\end{figure}
The EoS parameter for DE based on the constrained values of parameter $w_{0}$ is illustrated in figure $(\ref{fig:13})$ and $(\ref{fig:14})$. In model I, the EoS parameter's value are $\omega_{DE}=-0.807$ and $\omega_{DE}=-0.773$ for CC and joint estimates respectively. In model II, the present day EoS parameter values are $\omega_{DE}=-0.756$ and $\omega_{DE}=-0.884$ at $z=0$ for CC and joint estimates respectively.\\
The behavior of EoS parameter for DE illustrates that at present, these models will possess quintessence kind of dark energy. However, in model I, the $\omega_{DE}$ would cross phantom divide line leading to the quintom scenario, having phantom dark energy dominated evolution in future. In Model II, EoS parameter predicts the accelerated universe expansion at present era but, in future it predicts the decelerating universe expansion. It is an interesting aspect of the model. 

\subsection{Statefinder diagnostics}\label{sec:6.4}
\begin{figure}[!htb]
\captionsetup{skip=0.4\baselineskip,size=footnotesize}
   \begin{minipage}{0.40\textwidth}
     \centering
     \includegraphics[width=8.5 cm,height=7.5cm]{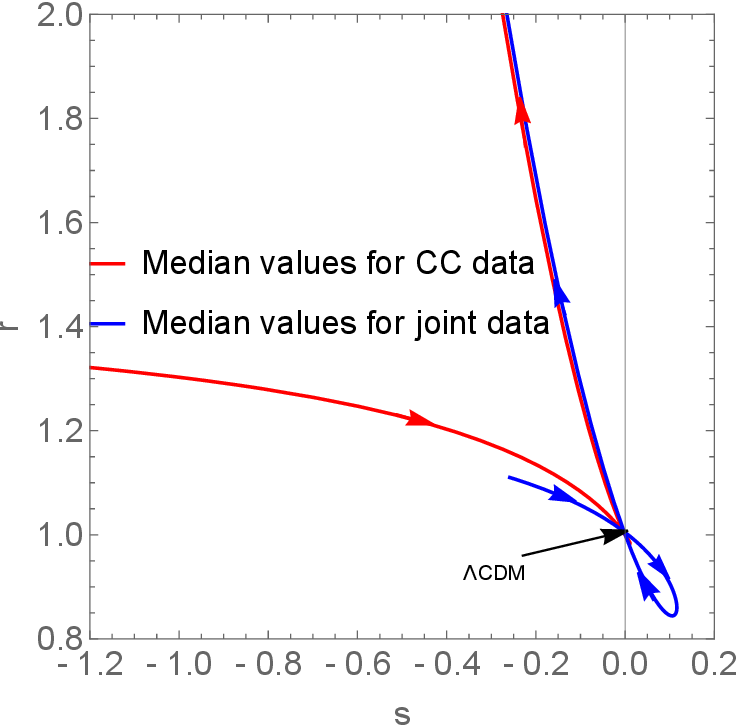}
\caption{Model I: $s$ and $r$ plane}
\label{fig:15}
    \end{minipage}\hfill
   \begin{minipage}{0.40\textwidth}
     \centering
     \includegraphics[width=8.5 cm,height=7.5cm]{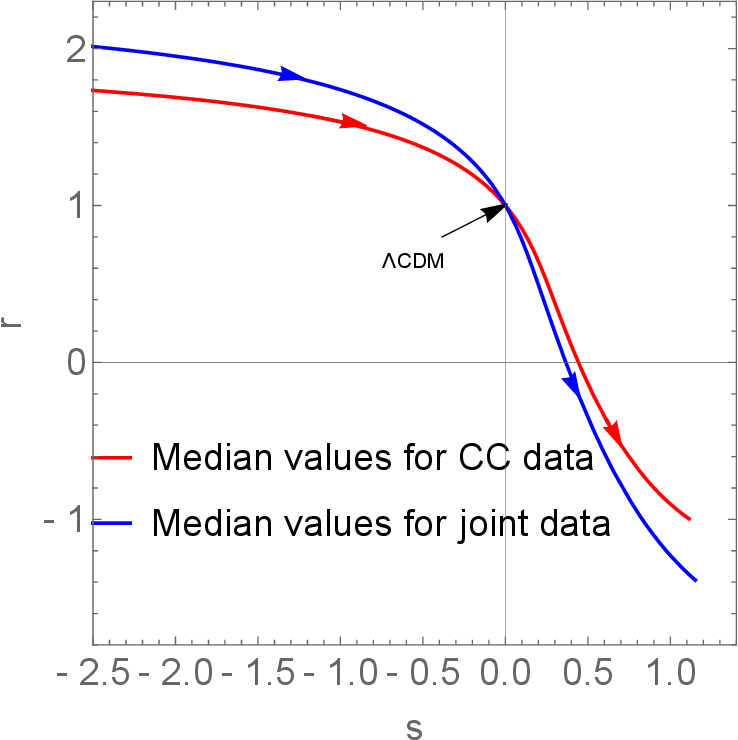}
  \caption{Model II: $s$ and $r$ plane}
\label{fig:16}
   \end{minipage}
\end{figure}

It is well recognized that geometric parameters may reveal the cosmological dynamics in the model. In order to find different dark energy models other than the $\Lambda$CDM model, it is necessary to investigate a few additional parameters besides $H$ and $q$. Therefore, the higher derivatives of $a(t)$ other than $q$ and $H$ may serve as an essential ingredients to characterize the dynamical nature of the cosmos. \\
In order to explain and differentiate the dynamics of various dark energy models, Sahni et al.\cite{sahni2003statefinder} proposed a couple of geometrical parameters $(r, s)$. This method, referred to statefinder diagnostic, is a helpful tool for understanding of the dark energy evolution. These statefinder diagnostic couple $(r, s)$ are described as
\begin{equation}{\label{109}}
r=\frac{\dddot a}{aH^{3}}, \  \ s=\frac{(r-1)}{3(q-0.5)} \  \  \  \text{where} \  \  \  \ q \neq 0.5
\end{equation}
The numerous DE models described in the literature can be expressed as follows for varying values of the statefinder pair $(r, s)$. The Chaplygin gas model would correspond to $r>1, s<0$ region of $r-s$ phase space. The $\Lambda$CDM model will have $r=1, s=0$ as the fixed point. The quintessence models would have $r<1, s>0$ region in phase space. For holographic dark energy models, $r=1, s=\frac{2}{3}$ point is a fixed point. The standard cold dark matter model will have $r=1, s=1$ in $r-s$ plane. \\
We exhibit the $r-s$ trajectories for the model parameters's median values in figure (\ref{fig:15}) and (\ref{fig:16}) for Model I and Model II respectively. Figure (\ref{fig:15}) demonstrates that for CC estimates, the progression of the statefinder pair $(r,s)$ corresponds to the Chaplygin gas models at early time and passes through the $\Lambda$CDM point and will finally evolve by unifying the dark matter and dark energy in the model. Figure $(\ref{fig:16})$ depicts that the progression of $(r,s)$ parameters starts like the Chaplygin gas models at early time and passes through $\Lambda$CDM point and finally evolve like the quintessence model during present era for both CC and joint data estimated values.

\subsection{The age of Universe}\label{sec:6.5}
In terms of $\mathit{z}$, the cosmic age $t(z)$ may be determined as ~\cite{tong2009cosmic}
\begin{equation}{\label{34}}
t(z)= \int_{z}^{\infty} \frac{dz}{H(z)(1+z)} \,dz. 
\end{equation}
We numerically calculate the universe's present age by determining the above integral for the Hubble parameters $H(z)$ given in Eqs. (\ref{23}) and (\ref{103}).\\
For Model I, the universe present age value is $t(z = 0) = 13.17$ Gyr and $t(z = 0) = 13.54$ Gyr for CC and joint estimates respectively which are close to the present age values of $\Lambda$CDM model obtained from the latest  observations \cite{2020A&A...641A...6P}. \\
For model II, the universe present age value is $t(z = 0) = 12.77$ Gyr and $t(z = 0) = 12.45$ Gyr for CC and joint estimates respectively. Clearly, the absence of structure formation era in this model alters the age of universe.

\section{Conclusions}\label{sec:7}
The dark energy equation of state having non-linear form may induce the non-metricity appearing in the symmetric teleparallel gravity. Under reasonable conditions, we study the universe evolution dynamics within the flat FLRW framework in symmetric teleparallel gravity termed as  $f(Q)$ gravity using the non-linear forms of dark energy EoS parameter. These parameters are commonly known as the Gong-Zhang parameterization and, are given by $ \omega_{DE}(z)=w_{0}(1+z)^{-1} $ and $\omega_{DE}(z)=w_{0}(1+z)^{-1} \exp\left(z/1+z\right)$. The dynamical evolution in these models are determined by using the derived Hubble parameter. The expansion rate of universe is described by this parameter and, one may also use it to  describe the non-metricity evolution in the $f(Q)$ gravity. The model parameters are constrained by using the Bayesian methods with MCMC analysis and the data of sneIa along with cosmic chronometer. We determine the median of posterior distribution subjected to the observations of cosmic chronometer and sneIa Pantheon data. \\
The summary of estimated parameters with cosmological parameters are given in Table \ref{table:1} and \ref{table:2} for both of these models. These models may explain the accelerating universe expansion of present era but the past and future dynamical evolution may be different in these models. It is an interesting finding of this study. These models are well behaved and observationally compatible but will have different evolution dynamics in far future.   \\
In the present study, we investigate the evolution dynamics using different cosmological parameters. The best fit curve of $H(z)$ reveals the compatibility of models with cosmic chronometer observations. The deceleration parameter behaviors of both the models highlight the evolution and dominating components during different epochs. The deceleration parameter's evolution curves demonstrate transition in the universe expansion from decelerated into accelerated phase. In model I, we obtain the present values of deceleration parameter as $q_{0}=-0.6170$ and $q_{0}=-0.5827$  for CC and joint estimated values. In model II, the present day values of deceleration parameter are $q_{0}=-0.2576$ and $q_{0}=-0.4120$ for CC and joint estimated values. This negative value of deceleration parameters at present ($z=0$) shows that the cosmos expansion is accelerating for both Hubble parameters. However, the past and future evolution scenarios in both of these models may be quite different.
In model I, one may have the cold dark matter dominated era in the past and in future, the universe will expand at super-exponential rate under the influence of phantom kind of dark energy. On the other hand, in model II, the cold dark matter dominated era is absent leading to the absence of structure formation era. And, this model will also possess the future deceleration era which may be possible under the energy exchange between dark energy and dark matter sectors. This is a peculiar difference between solutions for both of the considered models.\\
We also discuss about the physical parameters such as the energy density ($\rho_{DE}$), pressure ($p_{DE}$), and EoS parameter ($\omega_{DE}$). In this $f(Q)$ gravity model, the energy density for DE illustrates positive behavior but the pressure for DE may become negative from the recent past.  The cosmic cosmos may be expanding more quickly in this late era due to the pressure's negative nature. \\
The $\omega_{DE} $ based on the constrained values of parameter $w_{0}$ is illustrated in figure $(\ref{fig:13})$ and $(\ref{fig:14})$ for model I  and II respectively. In model I, at $z=0$, $\omega_{DE}=-0.807$ and $\omega_{DE}=-0.773$ for CC and joint data estimates. Similarly, for model II, $\omega_{DE}=-0.756$ and $\omega_{DE}=-0.884$ for CC and joint data estimates. During present times, all these values are pointing for the presence of quintessence kind of dark energy in models. Overall, at present times, we find that the $f(Q)$ model agrees with current observations of the accelerating cosmos. \\
In the $f(Q)$ gravity model, we estimate the present age of universe. In model I, the value is $t(z = 0) = 13.17^{+0.30}_{-0.19}$ Gyr  and $t(z = 0) = 13.54^{+0.02}_{-0.43}$ Gyr for CC and joint estimates respectively. On the other hand, in model II, the universe's present age value is $t(z = 0) = 12.77^{+0.10}_{-0.10}$ Gyr and $t(z = 0) = 12.45^{+0.07}_{-0.31}$ Gyr  for CC and joint estimates respectively. \\
A general reason for the different behavior of model II as compared to model I may be attributed to the term $\exp\left(\frac{z}{1+z}\right)$ involved in EoS parameter. This term alters the dynamics at high redshift during past as well as in the asymptotic era as $z\rightarrow -1$. Due to this term, the cold dark  matter dominated era does not exist in the model. As a consequence, the cosmological parameters such as deceleration parameter, age of the universe, EoS parameter of this model are quite different from the $\Lambda$CDM model as well as the model I governed by $\omega_{DE}=\frac{w_0}{1+z}$.
 
\section*{\textbf{Acknowledgements}}
We are grateful to the  honorable reviewer(s) for insightful comments on the manuscript. GPS and AS are thankful to the Inter-University Centre for Astronomy and Astrophysics (IUCAA), Pune, India for support under Visiting Associateship program. 


\end{document}